\documentclass[twocolumn]{article}   
\usepackage{times,epsfig,subfigure,calc,multicol,pslatex,comment}
\usepackage{cuted,url,textcomp,epstopdf,algorithmic,algorithm,color}
\usepackage{amsbsy,amsmath, amssymb,comment,dsfont,widetext,mathtools,multirow,amsthm}
\usepackage{framed}

\newtheorem{definition}{Definition}\newtheorem{theorem}{Theorem} \newtheorem{corollary}{Corollary}%\newtheorem{proof}{Proof} 
\newcommand{\bth}{\begin{theorem}}\newcommand{\ethe}{\end{theorem}} \newcommand{\bpr}{\begin{proof}}\newcommand{\epr}{\end{proof}} \newcommand{\ble}{\begin{lemma}}\newcommand{\ele}{\end{lemma}} \newcommand{\bco}{\begin{corollary}}\newcommand{\eco}{\end{corollary}}
\newcommand{\bde}{\begin{definition}}\newcommand{\ede}{\end{definition}}
\newcommand{\opl}{\oplus} \newcommand{\ba}{\wedge}      %\mathrel{\&}
\begin{document}
%	\begin{frontmatter}
\title{Secure Multi-Party Computation with a Helper\footnote{This work extends the conference paper\cite{sch15ab}.}}

\author{Johannes Schneider} %\footnote{Work conducted while affiliated with ABB Corporate Research, Baden-Daettwil, Switzerland.}}

%\institute{Johannes Schneider\at	University of Liechtenstein, Vaduz, Liechtenstein\\	\email{johannes.schneider@uni.li}  } %\author{Johannes Schneider} \institute{ ABB Research,Baden-Daettwil,Switzerland\\johannes.schneider@ch.abb.com}\maketitle
\maketitle% \normalsize \vfill
\begin{abstract}
A client wishes to outsource computation on confidential data to a network of parties. He does not trust a single party but believes that multiple parties do not collude. To solve this problem, we use the idea of treating one of the parties as a helper. A helper assists computation only. Often using more parties ensures confidentiality despite more corrupted parties. This does not hold for adding a helper. But a helper can in some cases lower the amount of communication asymptotically to the theoretical minimum of one bit per AND gate, improving significantly on schemes without a helper. It can also allow for very efficient computations of certain functions, as we show for the exponential function with public base.
\end{abstract}

%\begin{keywords}\\ \noindent Big data, client-server computation, secure cloud computing, secure multi-party computation, privacy preserving data mining
%\end{keywords}
%\end{frontmatter}
%\onecolumn

\section{Introduction}\label{sec:intro}
\begin{figure}[!htp]
	\centerline{\includegraphics[width=0.7\linewidth]{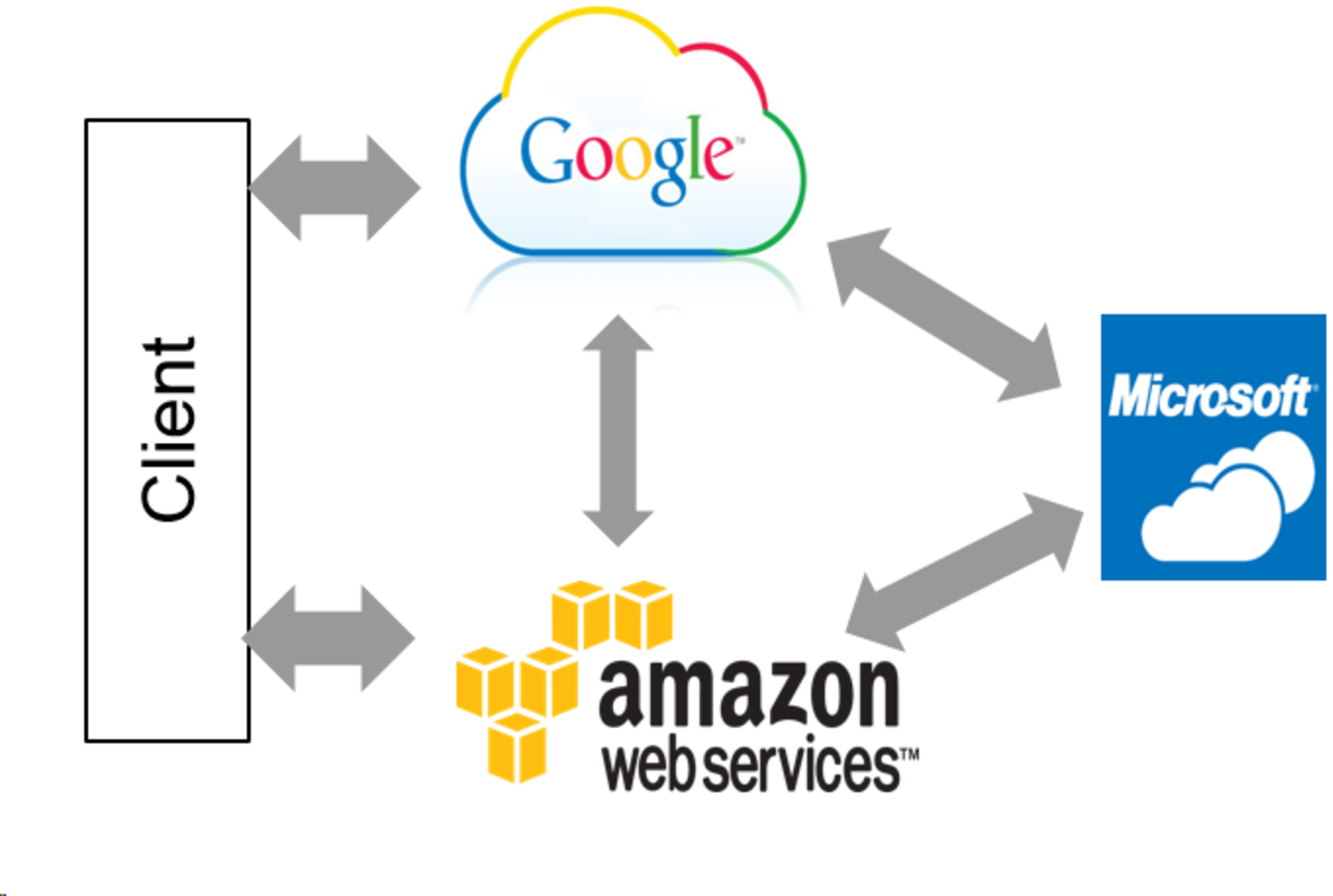}}
	\caption{A client outsources computation believing that no two cloud providers are dishonest. The right most provider serves as helper.}
	\label{fig:cloud}
\end{figure}

Cloud computing is on the rise with security remaining as one of the key challenges. A cloud provider must be fully trusted to refrain from misusing any confidential information. Often a single vulnerability, e.g. a corrupt system administrator, can put the confidentiality of a large amount of data at risk. Secure multi-party computation (MPC) is a computationally efficient approach that allows a client to outsource computation to a group of parties (or cloud providers), assuring that the client's information is prevented from misuse even if some parties are corrupted or cannot fully be trusted. Interaction is illustrated in Figure \ref{fig:cloud}, where a client uses various cloud providers and one of them serves as a helper to facilitate computation.
Dedicating one party for a special purpose, i.e. as a helper, is in contrast to existing schemes \cite{Gol87,bea90,ara12,bar89,ben88,bog08}. Typically shares of a secret are distributed equally among all parties, and all parties behave identically, i.e. they perform the same computations but on different values.  It is not clear, whether there is any benefit in deviating from this well-established body of work. Thus, the question we seek to answer is:
\smallskip\\
\emph{What are the advantages and disadvantages of using a helper, i.e. one party for assisting computation?}\smallskip 

\begin{table*}[!htp]
	\begin{center}
		\scalebox{0.85}{
			\begin{tabular}{ | l | l |l|l|l|} \hline
				\multirow{2}{*}{Paper} &\multicolumn{2}{c|}{Transmitted Bits for }	& Rounds & Maximum    \\ \cline{2-3} 			
				& AND of two single bits & AND of all pairs of $v$ variables && Corrupted Parties\\ \hline
				GMW '87\cite{Gol87} & $>$50 & $ > 3\cdot v^2$ & $2$& 2 \\\hline
				BMR '90\cite{bea90} & $>$10 & $ > 3\cdot v^2$& $>2$&2\\\hline
				CCS '16 \cite{ara12} & 3 &  $3\cdot v(v-1)/2$ &1&2\\\hline
				\textbf{JOS (This work)} & 5 & $\mathbf{v(v-1)/2 + 4v}$& 2&1\\\hline 
				%\textbf{JOS (This work)} & 7 & $v^2/2 + 4v$& 2&4\\\hline\hline 	
				%	& \multicolumn{2}{c|}{32 bit multiplications} & &\\ \hline
				%	Beaver '89\cite{bar89} & $>$300 &$> 3\cdot v^2$&  $>$2 &2  \\\hline
				%		BGW '88\cite{ben88},GRR '98 \cite{Gen98} & 390&$> 3\cdot v^2$&1& 2\\\hline			
				%		Maurer '06\cite{mau06} & 390 &$> 3\cdot v^2$ &1&2\\ \hline
				%		Sharemind '08\cite{bog08} & $\geq$ 500 &$> 3\cdot v^2$& 2&2 \\\hline
				%		CCS '16 \cite{ara12} & 96 &$96\cdot v(v-1)/2$ &1  &2\\\hline
				%		\textbf{JOS (This work)} & 160& $\mathbf{32\cdot v(v-1)/2 + 192v}$ &2&1 \\
				%\textbf{JOS (This work)} & 224& $32 v^2/2 + 224v$ &2&4 \\	\hline 	
			\end{tabular}
		}
	\end{center} 	
	%\vspace{-15pt}
	\caption{Comparison of three party protocols. XOR requires no communication in all schemes.}\label{tab:co}
\end{table*}
To this end, we develop techniques for MPC computation using a helper and compare them to the state-of-the-art in MPC focusing on the three party case. One might conjecture that the use of a helper does not help in improving security, since it does not hold a share of a secret. This is indeed one of our findings. One might also conjecture that the minimum number of rounds for any non-trivial operation is at least two, since the helper must receive information and return information. In fact, several of our protocols match this (lower) bound. With respect to other classical metrics such as communication and local computation complexity, it seems harder to come up with reasonable conjectures. \\
Our protocols using a helper are designed to minimize communication and round complexity, while keeping local computation complexity at the same level as the state-of-the-art. However, we also present a method involving a helper allowing to trade-off communication and round complexity for logical operations, i.e. large fan-in AND gates. Communication complexity has gained a lot in importance lately. All modern schemes for data analysis, such as Hadoop or Spark, rely typically on data-parallelism, i.e. performing the same computation for different parts of the data. Computation is done in a distributed fashion, which requires moving large amounts of data between computers (as is needed for MPC, in particular for big data). The bottleneck becomes bandwidth, i.e. the amount of information that can be transmitted, and not network latency, i.e. number of communication rounds. Since any Boolean circuit can be computed using basic building blocks such as `AND' and `XOR' gates, most MPC schemes focus on evaluating such circuits. Therefore, a natural question is: \smallskip \\ \emph{How many bits must be transmitted using a helper to evaluate AND and XOR gates with non-trivial security guarantees?} \smallskip

Intuitively, this seems to be at least two bits per (AND) gate, since a helper must receive and transmit at least one bit. However, surprisingly, in case the same variables occur multiple times, we can reach the theoretical minimum of just one bit per gate as indicated in Table \ref{tab:co} with non-trivial security guarantees under the commonly found assumption that randomness is pre-shared among parties. This is almost a factor 3 less than prior work for three party computation. Unfortunately, this comes at a price that anyone striving for performance might well be willing to pay. Traditionally, the entire system is corrupted if any confidential information becomes available to any party. Honest, non-corrupted parties are (implicitly) also assumed to abuse confidential information. We assume that the system is only compromised if a corrupted party, i.e. the attacker, gets secret information, but honest, non-corrupted parties participating in the computation would not abuse confidential information. Essentially, this means that an attacker can only corrupt one party in our scheme but two in other schemes as indicated in the last column of Table \ref{tab:co}. Arguably the largest benefits of using a helper can be reaped when computing special functions such as the exponential function in this work, where mathematical laws involving two variables can be used to compute the function in just two rounds.\\

To summarize, a helper can imply a significant reduction of communication complexity in several situations under the condition that the system is only compromised if a corrupted party, i.e. the attacker, gets secret information.
\begin{itemize}
	\item  If the same variables occur in multiple operations so that the effort of sharing with the helper can be reduced.
	\item  If the function to be computed allows leveraging the secret sharing principle of having encrypted values and keys. %This is illustrated for the computation of the exponential function in this work, where mathematical laws involving two variables can be used to compute the function in just two rounds. % prior work \cite{sch15b}
\end{itemize}

\subsection{Contributions}
\noindent In summary, we make the following contributions:

\begin{itemize}
	\item We elaborate on the ``distrust attacker'' model assuming semi-honest parties and provide a mapping of security guarantees to the standard ``distrust all'' security model.
	
	\item We assess the usefulness of a helper assisting computation. We present a new scheme for MPC in the semi-honest model for boolean gates using three parties with non-trivial security guarantees. If variables occur multiple times, we improve on the amount of communication needed to compute multiple ANDs compared to prior work (see Table \ref{tab:co} and Related Work Section). We even reach an asymptotically optimal value of just one bit per AND gate if the number of evaluated AND gates is more than linear in the number of variables.
	\item We present a method to compute unbounded fan-in gates in constant rounds. It comes with a trade-off for communication and rounds. Using $w$ variables and messages of size O($w\cdot 2^{w-k}$) for arbitrary $k \in [2,w]$ an AND can be computed in O($\log k$) rounds and O($w\cdot 2^{w-k}$) operations involving single bits. 
	\item We present a statistically secure protocol for computing any exponential function $a^x$ for a secret exponent $x$ and public base $a$ improving prior work (even in the same model) considerably. %This adds to prior work on \cite{sch15b}.
	
\end{itemize}
\subsection{Outline}
After stating the model in Section \ref{sec:mod}, we provide an overview of the method and its motivation in Section \ref{sec:over}, followed by describing the secret sharing in Section \ref{sec:enc} and basic operations in Section \ref{sec:addXOR}, i.e. XOR and NOT. The basic ideas of the AND protocol are stated in Section \ref{sec:homo} using four parties, i.e. two helpers. The number of helpers is reduced to one in Section \ref{sec:3par}. In Section \ref{sec:mulAnd} we derive a protocol that achieves amortized communication of just 1 bit per AND gate, if the same variables occur in a large number of gates. A trade-off between communication and round complexity for arbitrary fan-in gates is given in Section \ref{sec:arb}. A protocol for computing exponential functions with a public base is given in Section \ref{sec:exp}. % We extend to more than three parties using just one helper in Section \ref{sec:Opt}. \\

\section{Model}\label{sec:mod}
We adopt the semi-honest model for client-server computation, where a curious but passive attacker can monitor a party completely, e.g. its memory, disc and CPU registers. A client holds an arbitrary amount of secret values. The client wishes to evaluate a function using three parties, such that no corrupted party, i.e., attacker, can learn anything about the input or the output. The standard model is different since it is based on distrusting everybody. This means that no party should learn anything about the input.  The standard, i.e. ``distrust all'', model assumes that honest, non-corrupted parties do not actively share any information with any other party throughout computation but can still not be trusted with any confidential information, i.e.,. they behave dishonestly as soon as they get secrets.  In our model, an honest party would not misuse secret information, but an attacker would. We call this ``distrust attacker'' model. Any result for a ``distrust all'' model can be directly translated to the ``distrust attacker'' model: % In the Oxford dictionary, ``honest'' is defined as ``free of deceit; truthful and sincere''. In that sense, our model assumes that parties do behave honestly also when being given confidential information.

\bth
An MPC scheme remains confidentiality of secrets despite $x$ corrupted parties for a ``distrust all'' model, if and only if it remains confidentiality despite $x+1$ corrupted parties in the ``distrust attacker'' model.
\ethe
\bpr
Assume that corrupted parties share their information with all parties.  Assume that for a set of $x+1$ corrupted parties $S$ in the distrust all model, at least one party $A$ (not necessarily corrupted) can obtain some confidential information using the information from the parties $S$. Therefore, in the distrust attacker model, assume that $S \cup \{A\}$ are corrupted. In this case, all corrupted parties, i.e., also the attacker, obtain all information from the $S \cup \{A\}$ parties and, thus, also confidential information. Therefore, if an MPC schemes withstands $x$ corrupted parties in the distrust all model it cannot withstand more than $x+1$ parties in the distrust attacker model.\\
Assume that for $x+2$ corrupted parties $S$ in the distrust attacker model, at least one corrupted party $A \in S$ can obtain some confidential information. Therefore, in the distrust all model, assume that $S \setminus A$ are corrupted. Since $A$ receives by assumption all information from parties $S\setminus A$ it obtains confidential information.  
\epr
In all MPC schemes, typically, a client encrypts its secrets and distributes the shares among the parties, the parties compute the desired function and return their shares to the client. The client itself does not participate in the computation and, therefore, does not count as a party. This differs from the classical MPC model, where each party holds a secret (or at least a share), and the output should be known by (at least) one party. We can emulate the classical model: The parties can always obtain the secret value of an output through collusion (rather than transmitting all their shares to the client). In case each party has a secret, each party can execute the same protocol for encryption and distribution of shares of its secret as a client having all secrets. Thus, the extension to using several clients (each having some secret value) is obvious.

Our simplest network consists of a client, a key holder (KH) and an encrypted value holder (EVH) and a helper. The client communicates with the KH and EVH.  A network with three parties is shown in Figure \ref{fig:cloud}. We assume perfectly secure communication channels between parties. We assume that there is pre-shared randomness among pairs of parties. This implies that keys do not have to be transmitted but can be regareded as pre-shared. In practice, one can generate keys using a shared secret seed and pseudo-random number generators. %Assume that 

\section{Overview of Approach} \label{sec:over}
Our scheme called \emph{JOS} requires at least three parties, where each party can be thought of having a dedicated role: a keyholder (KH), an encrypted value holder (EVH) and a helper. Beyond three parties, the number of keyholders increases. Generally, the keyholder stores keys but it has no access to ciphertexts. The encrypted value holder stores encrypted values but no keys. Note, there are circumstances, where this distinction has to be seen less strict, e.g., there might be two encryptions of the same secret such that one is held by the KH.  A helper assists computations. It might obtain keys and encrypted values that do not match, i.e. none of its keys can be used to decrypt any of its encrypted values.  The helper should only facilitate computation, which is a key conceptual feature of our method. For example, assume the parties should store a large amount of data. Since all other methods require shares (of at least the same size as ours) being held by all parties, each party must store its shares somewhere, whereas in our case only the EVH and the KH need to store data. Thus, we improve on the amount of storage needed for three party protocols and match those of two party protocols.\\

The (mathematical) motivation to use helpers is that the AND of two numbers can be computed by combining four parts consisting only of encrypted values and keys. In a naive computation, each part can be computed by one party, encrypted and combined by the KH and EVH to yield an encrypted AND of the two numbers. This would yield a total of four parties including two helpers. Through double encryption of values, we can reduce the number of parties to three. The derivation of the protocols uses the associative and distributive properties of linear secret sharing. Our scheme is illustrated for three encryption schemes based on XOR and addition. The given protocols perform efficient Boolean operations (AND, XOR) operations. They could also be extended to arithmetic operations (addition, multiplication). Linear secret sharing strikes through little computational overhead -- in particular when compared to protocols that require cryptographic primitives, e.g. generation of prime numbers.

\section{Encryption}\label{sec:enc}
We use linear secret sharing, but only two out of three parties obtain a share, i.e. in most situations we generate only two shares. We label these shares differently, i.e. we have one encrypted value and, potentially, multiple keys. The distinction between keys and encrypted values is sometimes helpful, e.g. for computing an exponential function (Section \ref{sec:exp}). %\\%or the sine function \cite{sch15b}.\\
%\subsection{Encryption for XOR and AND} \label{sec:bitop}
%\noindent\underline{Encryption for Boolean Operations:}
For a given bit $m\in \{0,1\}$ we choose a random key $K \in \{0,1\}$. The encryption $ENC_K(m)$ of bit $m \in \{0,1\}$ using key $K$ is the XOR ($\opl$ symbol), i.e. $ENC_K(m):=K \opl m$.  The decryption $DEC_K(c)$ of a ciphertext $c$ is $DEC_K(c):=c \opl K$.\\
%\noindent\underline{Encryption for Arithmetic Operations:}
%For a given number $m\in \{0,1\}^{l}$ of $l$ bits we choose a random key $K \in \{0,1\}^b$ using $b\geq l$ bits.  The encryption $ENC_K(m)$ of number $m$ using key $K$ is then $ENC_K(m):=(K+m) \mod 2^b$.  We denote inverse elements with superscripts, e.g. the inverse of $K$ is $K^{-1}:=2^b-K$. The decryption $DEC_K(c)$ of a ciphertext $c$ is $DEC_K(c):=(c+K^{-1}) \mod 2^b$. % The decryption of an encrypted message yields the original message, i.e. , $DEC_K(ENC_K(m)):= (K^{-1}+(m+K \mod 2^b)) \mod 2^b = m$.
%Whereas the above encryption in an arithmetic ring guarantees unconditional security, we also mention a variant without modulo that is only statistically secure. If we just add the key, we have: $ENC_K(m):=K+m$, $DEC_K(ENC_K(m))=(m+K) +K^{-1}  = m+K-K = m$.

\section{XOR and NOT Operations} \label{sec:addXOR}
We discuss the entire process ranging from encryption of plaintexts to decryption of results for a single operation. To compute a (bitwise) XOR of two numbers $a,b$ the client encrypts both $a$ and $b$. It sends the two keys to the KH and the encrypted values to the EVH. As a next step, the KH computes the XOR of the two keys and the EVH the XOR of the two encrypted values. Both send their results back to the client. The client obtains $a \opl b$ by decrypting the result from the EVH with the key received from the KH.\\

A NOT operation (denoted by $\neg$) corresponds to computing an XOR of an expression and the constant one. It can be done by the EVH by computing the `NOT' of the encrypted value. More mathematically, we have $\neg a = a \opl 1$ and $\neg ENC_{{K_a}}(a) = \neg (a \opl K_a) = (\neg a) \opl K_a = ENC_{{K_a}}(\neg a) $.

\section{Basic Ideas for an AND Gate} \label{sec:homo}
To illustrate the main ideas, we discuss the AND gate for two numbers using two helpers aside from the EVH and the KH. Later, we refine the algorithms to require only one helper, i.e. three parties, and extend this to multiple parties. Note, all our main results are based on the three party protocols using one helper. The protocol could be extended to multiplication working in an analogous manner. The key idea is to express the AND of the plaintexts using several parts, each consisting of an encrypted value and a key. Each part can be computed on a separate party, e.g. we use (and prove) that
\begin{flalign} \label{eqM}
a \ba b & =  \big(ENC_{K_a}(a) \ba ENC_{K_b}(b)\big)  \opl ({K_a}  \ba {K_b}) \nonumber \\
& \opl \big({K_a} \ba ENC_{K_b}(b)\big)  \opl \big({K_b} \ba ENC_{K_a}(a)\big)
\end{flalign}

\begin{figure*}[!htp]
	\centerline{\includegraphics[width=0.9\linewidth]{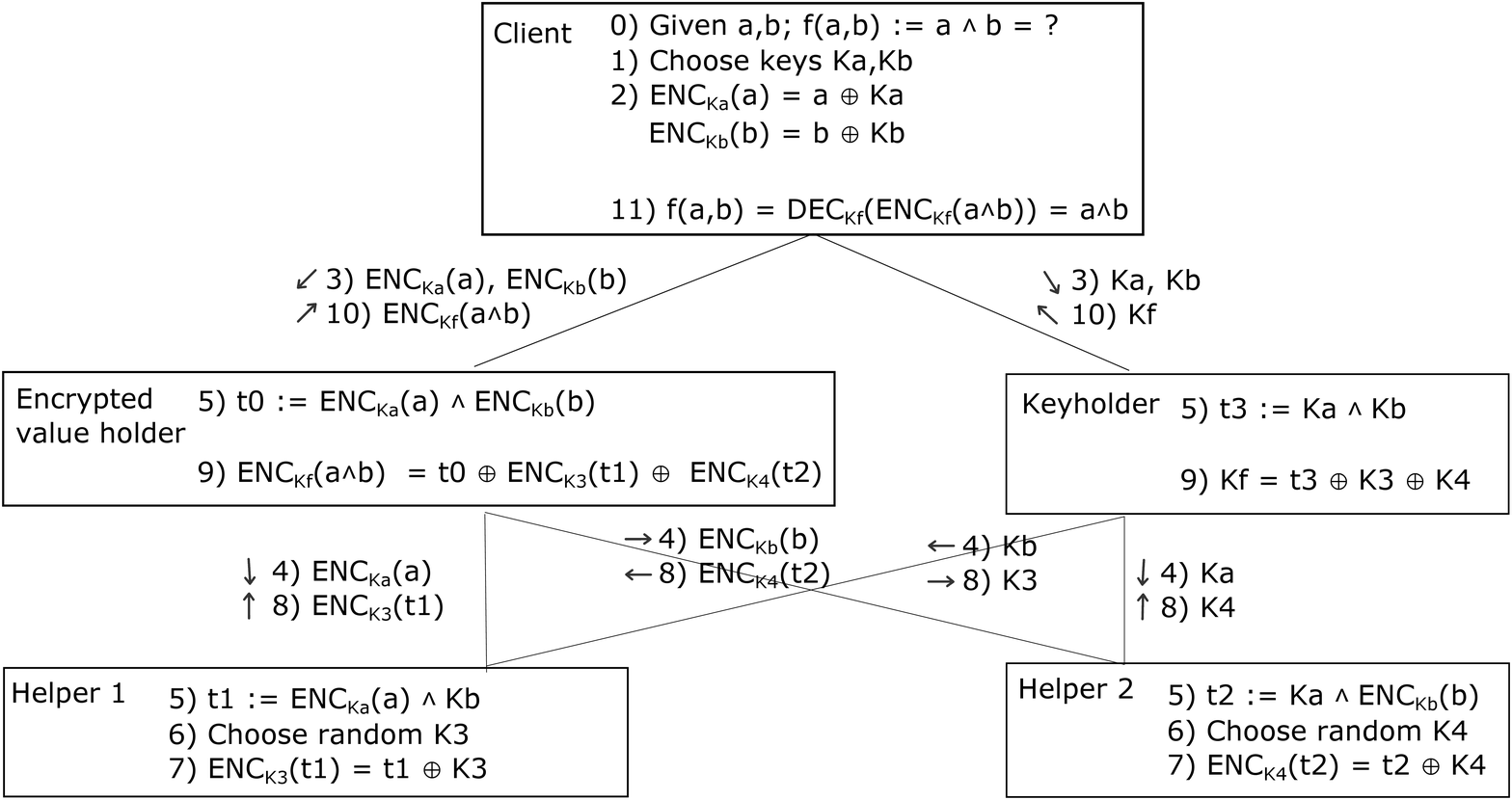}}
	\caption{For comprehension purposes: Algorithm for an AND ($\ba$) of two bits using two helpers. Later protocols use one helper only.}
	\label{fig:and}
\end{figure*}

Each of the four terms $ENC_{K_a}(a) \ba ENC_{K_b}(b)$, ${K_a} \ba ENC_{K_b}(b)$, ${K_b} \ba ENC_{K_a}(a)$ and ${K_a}  \ba {K_b}$ is computed by one party. Each of the two helpers chooses a key to encrypt its part before sharing the encrypted part with the EVH and the key with the KH. The KH and EVH combine all partial results to obtain the key and encrypted value of $a \ba b$. The algorithm to compute the AND of two bits is shown in Figure \ref{fig:and}.
In Steps 1 to 3, the client prepares the computation. Step 4 can also be seen as part of the preparation, i.e. secret sharing. The actual computation of the AND consists only of steps 5-9.  No keys must be transmitted during computation if keys are pre-shared or randomness is created using a shared secret seed and pseudo-random number generators (as we stated in the Model Section). 

Next, we prove that the protocol is secure and correct. Security can also be shown - it follows since no party can reveal any information about a secret by any combination of the values it has obtained. We prove it formally only for our best protocols, e.g. for the three party protocol in Section \ref{sec:3par}.

\bth \label{thm:and4}
The AND protocol in Figure \ref{fig:and} is correct.
\ethe

\bpr
To show correctness, i.e. that we indeed compute $a \ba b$, we must prove that the decryption done by the client yields the correct result. From Figure \ref{fig:and} we see that the final key delivered to the client is  $K_f:=t_3 \opl K_3 \opl K_4$. Thus, it remains to show that for the EVH holds the claimed equation in Step 9:
\begin{flalign} \label{eq0}
ENC_{K_f}(a\ba b) &= ENC_{t_3 \opl K_3 \opl K_4}(a\ba b) \nonumber
\\ & =t_0 \opl ENC_{K_3}(t_1) \opl ENC_{K_4}(t_2)
\end{flalign}

%\big(ENC_{{K_a}}(a)\ba ENC_{{K_b}}(b)\big) \opl \\ ENC_{K_3}({K_a}\ba ENC_{{K_b}}(b))\opl \\ ENC_{K_4}(ENC_{{K_a}}(a)\ba {K_b})  \opl \\ ENC_{K_2}({K_a}\ba {K_b})\\

We prove Equation (\ref{eqM}) first. It can be derived using basic laws such as distributiveness and associativeness, $x \opl x = 0$ and $x \opl 0 = x$ and, thus, $ a \opl x \opl x = a$:
%\begin{multline}
%\small{
\begin{flalign} \label{eq6}
a \ba b =&  (a \opl K_a \opl K_a) \ba b \nonumber\\ \nonumber
=& \big((a \opl K_a) \ba b\big)   \opl  \big(K_a \ba b\big)\\ \nonumber
=& \big((a \opl K_a) \ba (b \opl K_b \opl K_b)\big)  %\\ \nonumber
\opl  \big(K_a \ba (b \opl K_b \opl K_b)\big) \\ \nonumber
=& \big((a \opl K_a) \ba (b \opl K_b )\big)  %\\ \nonumber
\opl \big((a \opl K_a) \ba  K_b\big)  \\
&\opl  \big(K_a \ba (b \opl K_b )\big) \opl  \big(K_a \ba K_b \big)
\end{flalign}
%}
%\end{multline}
Note, by using the definition of the encryption $ENC_K(m)=m \opl K$ we obtain Equation (\ref{eqM}) from (\ref{eq6}).
Next, we prove the Equation (\ref{eq0}) starting from $ENC_{K_f}(a \ba b)$ substituting the final key $K_f:=t_3 \opl K_3 \opl K_4$:
\small{
	\begin{flalign}
	&ENC_{K_f}(a\ba b)  \nonumber \\
	=& ENC_{t_3 \opl K_3 \opl K_4}(a\ba b)  \text{\footnotesize{ \emph{( Step 9, KH, Figure \ref{fig:and})}}}  \nonumber
	\\=&(a \ba b) \opl (t_3 \opl K_3 \opl K_4) \nonumber
	\\ = &(a \ba b) \opl (K_a \ba K_b) \opl K_3 \opl K_4 \nonumber  \text{\footnotesize{ \emph{(Using $t_3:=K_a \ba K_b$)}} }
	\\ = & \big((a \opl {K_a}) \ba (b \opl {K_b})\big)   \opl \big({K_a} \ba (b \opl {K_b})\big) \nonumber
	\\ & \opl \big({K_b} \ba (a \opl {K_a})\big)  \opl (K_a \ba K_b)  \opl (K_a \ba K_b) \nonumber
	\\  &  \opl  K_3 \opl K_4  \nonumber \text{ \emph{(Using Eq. (\ref{eq6}))} }
	%	\end{flalign}
	%	\begin{flalign}
	\\= & \big((a \opl {K_a}) \ba (b \opl {K_b})\big)    \opl \big({K_a} \ba (b \opl {K_b})\big) \opl K_4 \nonumber
	\\  &\opl  \big({K_b} \ba (a \opl {K_a})\big) \opl K_3  \nonumber \text{ \emph{(Using $a \opl (x \opl x) = a$)} } \nonumber
	\\  = & \text{\footnotesize{$\big(ENC_{{K_a}}(a)\ba ENC_{{K_b}}(b)\big) \opl  ENC_{K_4}({K_a}\ba ENC_{{K_b}}(b))$}}   \nonumber
	\\ &\opl  ENC_{K_3}(ENC_{{K_a}}(a)\ba {K_b}) \text{ \footnotesize{\emph{(Using $ENC_K(m):= m \opl K$)}}}\nonumber
	\\   = &t_0 \opl ENC_{K_3}(t_1) \opl ENC_{K_4}(t_2) \nonumber
	\end{flalign}}
\epr

\begin{figure*}[h!tp]
	\centerline{\includegraphics[width=1.0\linewidth]{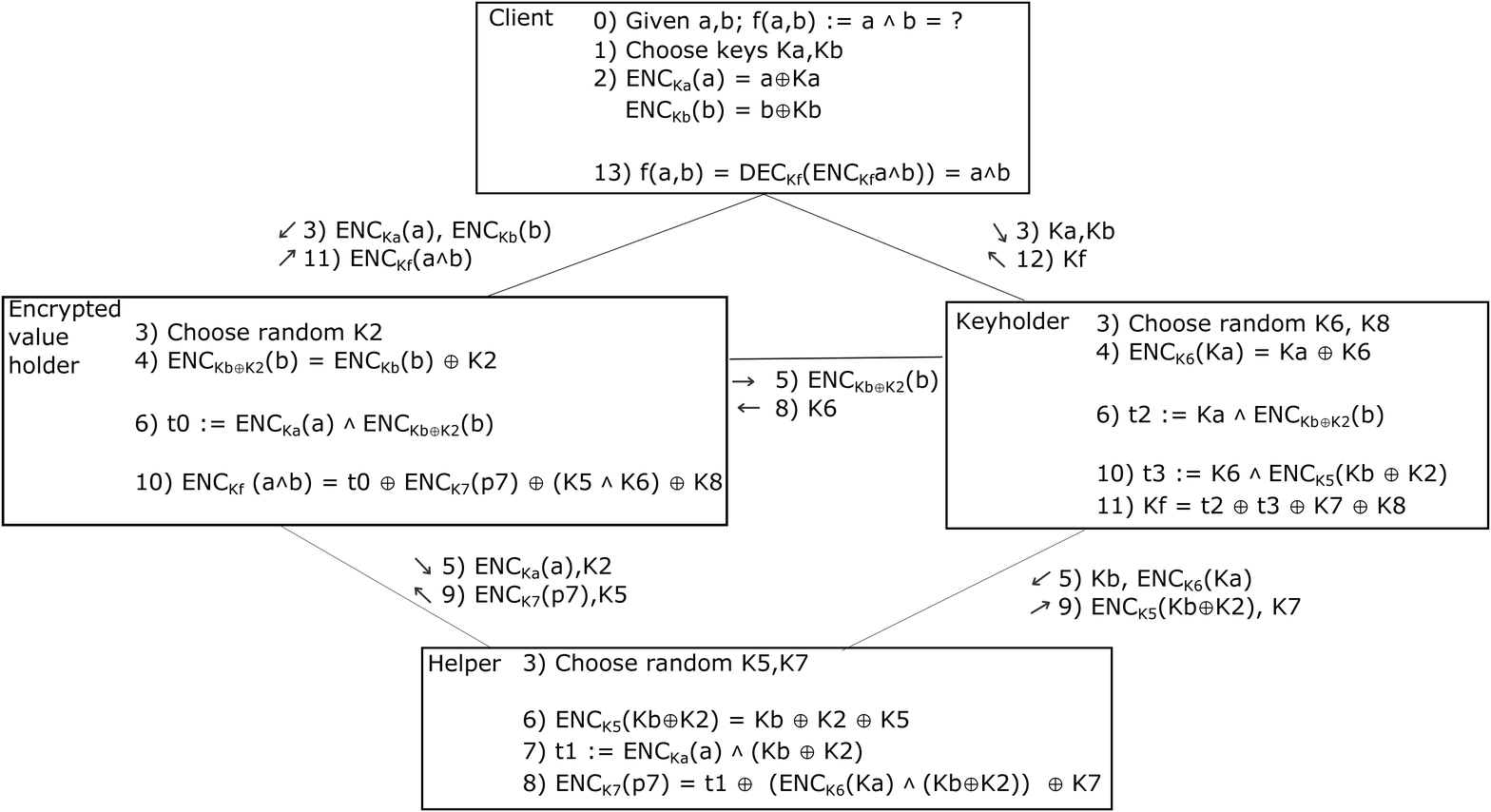}}
	\caption{Algorithm for an AND ($\ba$) operation of two bits using three parties.}
	\label{fig:3and}
\end{figure*}

\begin{comment}
\begin{figure*}%[!htp]
\centerline{\includegraphics[width=1.05\linewidth]{ThreePartyAndKeyHoldMulMod4.eps}}
\caption{Algorithm for multiplication of two integers using three parties.}
\label{fig:3mul}
\end{figure*}
\end{comment}

\section{Three Parties} \label{sec:3par}
Here, we reduce the number of parties from four to three but still focus on the computation of a single `AND'.
It is possible to use only one helper, i.e., three parties. We remove Helper 2. For an AND of two secrets $a,b$ encrypted with $K_a$ and $K_b$ Helper 2 computes $ENC_{K_b}(b) \wedge K_a$ (see Figure \ref{fig:and}). None of the other three parties can compute this expression using both $ENC_{K_b}(b)$ and $K_a$, since all parties hold at least one of the two values $K_b, ENC_{K_a}(a)$ and therefore they could reveal a secret.
However, the remaining helper (i.e. Helper 1) can compute on encrypted values, i.e. the EVH can encrypt $ENC_{K_b}(b)$ with a randomly chosen key $K_2$ to obtain a `double' encrypted value of $b$, i.e. $ENC_{K_b\opl K_2}(b)$. The helper can use $ENC_{K_b\opl K_2}(b)$ instead of  $ENC_{K_b}(b)$.  In particular, the EVH can double encrypt $b$ and it can share $ENC_{K_b\opl K_2}(b)$ with the KH (as long as the KH does not obtain $K_2$) and the key $K_2$ with the helper. This allows the KH to compute $K_a \wedge ENC_{K_b\opl K_2}(b)$, leaving to compute $K_a \wedge (K_b\opl K_2)$: We encrypt both values, i.e. $K_a$ with $K_6$ and $K_b \opl K_2$ with $K_5$ and compute using Equation (\ref{eq6}):
%\vspace{-6pt}
%\small{
\begin{flalign} \label{eq11}
&K_a\wedge (K_b \opl K_2) = \\ \nonumber
& ENC_{K_6}(K_a)\wedge ENC_{K_5}(K_b \opl K_2)  %\\ \nonumber
\opl K_5 \wedge ENC_{K_6}(K_a) \\ \nonumber
& \opl K_6 \wedge ENC_{K_5}(K_b \opl K_2) \opl (K_5\wedge K_6) 
\end{flalign}
%}
In this case, we do not need to distribute all four terms to (four) different parties.  In our scenario the (remaining) helper holds $K_b \opl K_2$, $K_5$ and $ENC_{K_6}(K_a)$. Thus, it can compute two terms, namely $ENC_{K_6}(K_a)\wedge ENC_{K_5}(K_b \opl K_2)$ and $K_5 \wedge ENC_{K_6}(K_a)$. We can even simplify for the helper: $ENC_{K_6}(K_a)\wedge ENC_{K_5}(K_b \opl K_2) \opl K_5 \wedge ENC_{K_6}(K_a) = ENC_{K_6}(K_a)\wedge (K_b \opl K_2)$.

This idea is realized in the protocol shown in Figure \ref{fig:3and}. The message complexity can be reduced by pre-sharing of keys. The very last key ($K_8$) is only needed if the result is used in further computations, e.g., we compute $(a \ba b) \ba c$ and reuse $(a \ba b)$.

Next, we show correctness and security of the AND protocol.
\bth \label{thm:3prt}
The AND protocol in Figure \ref{fig:3and} is  correct.
\ethe

\bpr
Analogously to the proof of Theorem \ref{thm:and4}, we show that decrypting the encrypted result (Step 10 for the EVH) with the final key (Step 11 for KH) in Figure \ref{fig:3and} gives $a \ba b$. We start by transforming $a\wedge b$ as shown below:% (see separate box).
\small{
	%\begin{strip}
	%\begin{framed}
	%	(Part of Proof of Theorem \ref{thm:3prt}.)
	%\begin{widetext}
	\begin{flalign*}
	a \ba b  = & \big((a \opl K_a) \ba (b \opl (K_b \opl K_2) )\big) \nonumber
	\opl \big((a \opl K_a) \ba  (K_b \opl K_2)\big)  \opl   \\ \nonumber
	&\big(K_a \ba (b \opl (K_b \opl K_2) )\big)\opl  \big(K_a \ba (K_b \opl K_2) \big)  \text{ \emph{(Using Eq. \ref{eq6})} }
	\\ = &\big((a \opl K_a) \ba (b \opl (K_b \opl K_2) )\big)
	\opl \big((a \opl K_a) \ba  (K_b \opl K_2)\big) \nonumber \\
	& \opl  \big(K_a \ba (b \opl (K_b \opl K_2) )\big) \nonumber \\ \nonumber
	&\opl  ENC_{K_6}(K_a)\wedge ENC_{K_5}(K_b \opl K_2) \opl K_5 \wedge ENC_{K_6}(K_a) \\ \nonumber
	& \opl K_6 \wedge ENC_{K_5}(K_b \opl K_2) \opl (K_5\wedge K_6)   \text{ \emph{(Using Eq. \ref{eq11})} }
	\\ = &  ENC_{K_a}(a) \ba ENC_{K_b \opl K_2} (b) \nonumber    \text{\emph{(Using $ENC_K(m) = K \opl m$)}}\\
	& \opl ENC_{K_a}(a) \ba  (K_b \opl K_2) \opl  K_a \ba ENC_{K_b \opl K_2}(b) \nonumber \\
	&\opl  ENC_{K_6}(K_a)\wedge ENC_{K_5}(K_b \opl K_2)\\ \nonumber
	& \opl K_5 \wedge ENC_{K_6}(K_a) \opl K_6 \wedge ENC_{K_5}(K_b \opl K_2) \opl (K_5\wedge K_6)\\ \nonumber
	= & t_0   \nonumber \text{ \emph{(Using $t_0:=ENC_{K_a}(a) \ba ENC_{K_b \opl K_2}(b)$}} \\ \nonumber
	& \opl t_1  \nonumber \text{ \emph{(Using $t_1:=ENC_{K_a}(a) \ba  (K_b \opl K_2)$)}} \\ \nonumber
	&\opl t_2 \text{ \emph{(Using $t_2:=K_a \ba ENC_{K_b \opl K_2}(b))$}} \\ \nonumber
	&\opl ENC_{K_6}(K_a)\wedge ENC_{K_5}(K_b \opl K_2) \\ \nonumber
	&\opl K_5 \wedge ENC_{K_6}(K_a)
	\opl K_6 \wedge ENC_{K_5}(K_b \opl K_2) \opl (K_5\wedge K_6)
	%\end{flalign*}	\begin{flalign*}
	\\ = & t_0 \opl (K_5\wedge K_6)
	\opl t_1  \opl K_7 \opl K_7  \nonumber \\ \nonumber
	& \opl ENC_{K_6}(K_a)\wedge ENC_{K_5}(K_b \opl K_2)\opl K_5 \wedge ENC_{K_6}(K_a) \\
	& \opl t_2 \opl K_6 \wedge ENC_{K_5}(K_b \opl K_2)  \nonumber  \\ \nonumber
	& \text{ \emph{(Rearranging and XOR with $K_7 \opl K_7=0$)} }
	\\ = & t_0 \opl (K_5\wedge K_6) \opl t_1  \opl K_7 \opl K_7
	\opl ENC_{K_6}(K_a)\wedge (K_b \opl K_2) \\
	&\opl t_2 \opl t_3  \nonumber \text{ \emph{(Simplifying and $t_3 :=  K_6 \wedge ENC_{K_5}(K_b \opl K_2)$)} }
	\\ = & t_0 \opl (K_5\wedge K_6)  \nonumber
	\opl t_1  \opl ENC_{K_6}(K_a)\wedge (K_b \opl K_2) \opl K_7 \\ \nonumber
	& \opl t_2   \opl t_3 \opl K_7
	\\ = & t_0 \opl (K_5\wedge K_6)
	\opl ENC_{K_7}(t_4) \opl K_7
	\opl \big(t_2   \opl t_3 \opl K_7\big)  \\ \nonumber
	&\opl K_8  \opl K_8  \nonumber\\ &\nonumber \text{ {\emph{(Using Def. of $ENC_{K_7}(t_4)$ in Step 8, Figure \ref{fig:3and})}}} %=t_1 \opl ENC_{K_6}(K_a)\wedge ENC_{K_5}(K_b \opl K_2) \opl K_7$
	\\ = & t_0 \opl (K_5\wedge K_6)
	\opl ENC_{K_7}(t_4) \opl K_7  \opl K_8 \opl K_f \nonumber  \\
	\nonumber & \text{ \emph{(Using $K_f=t_2   \opl t_3 \opl K_7 \opl K_8 $ in Step 10, Figure \ref{fig:3and})}}
	\\ = & ENC_{K_f}(a\ba b) \opl K_f = a \ba b \text{ \emph{(Using Step 10, Figure \ref{fig:3and})}} \nonumber
	\end{flalign*}
	%\end{widetext}
	%\end{framed}
	%\end{strip}
}
\epr

\bth \label{thm:3prtsc}
The AND protocol  in Figure \ref{fig:and}  is perfectly secure.
\ethe
Security follows since no party can reveal a secret by arbitrary combination of the values it has.
\bpr
We show that none of the parties can obtain any information about $a$ or $b$.\\
The EVH obtains keys $K_2,K_3,K_5, K_6, K_8$ and encrypted values $ENC_{K_a}(a),ENC_{K_b}(b), ENC_{K_7}(t_7)$. To get information about $a$ or $b$ the EVH needed to decrypt one of the encrypted values, i.e. it needed $K_a$ or $K_b$ or $K_7$. However, it has no information about these keys.
The KH obtains $K_a,K_b, K_6,K_7,K_8,ENC_{K_2 \opl K_b}(b)= ENC_{K_2}(b \opl K_b), ENC_{K_5}(K_b \opl K_2)$. The KH has neither $K_5$ nor $K_2$ so it cannot disclose any information about $a$ or $b$.
The helper obtains $K_b,K_2,K_5,K_7$ and $ENC_{K_a}(a),ENC_{K_6}(K_a)$. Since the helper has no information about $K_a$ and $K_6$, it cannot learn anything about $a$ or $b$.
\epr

\bth \label{thm:3bits}
The computation of the AND protocol (Steps 4-10) in Figure \ref{fig:and} needs a total of 5 transmitted bits.
\ethe
\bpr
In our model we assume that keys are pre-shared. The distribution of the secrets by the client is not part of the computation. Thus, only transmissions of encrypted values of Steps 5 to 9 are relevant, which yields a total of five bits, i.e. $ENC_{K_a}(a)$,$ENC_{K_b \opl K_2}(b)$, $ENC_{K_6}(K_a)$,$ENC_{K_7}(t_7)$ and $ENC_{K_5}(K_b \opl K_2)$.
\epr

\subsection{Multiple ANDs: Reusing Variables and Multiple Encryptions} \label{sec:mulAnd}
\begin{figure*}[!htp]
	\centerline{\includegraphics[width=1.05\linewidth]{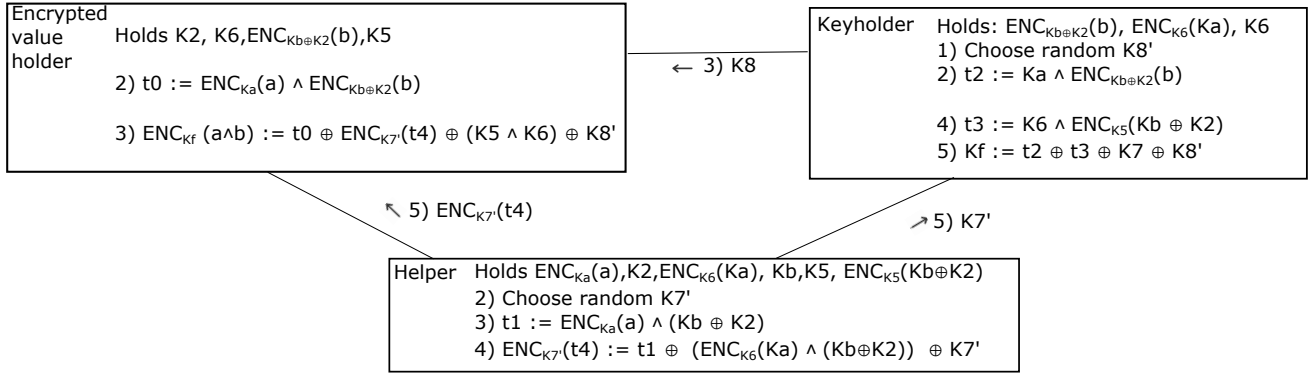}}
	\caption{Algorithm for an AND ($\ba$) operation of two bits $a,b$  reusing priorly shared values with the helper.}
	\label{fig:3andBoth}
\end{figure*}

We have shown that a single AND operation is perfectly secure, underlying the assumption that no party has both an encrypted value and a matching key. However, when using the same variables in multiple operations (but in different pairings), the amortized communication per gate is reduced, since the KH and EVH only need to share some terms with the helper once. For example, to compute $a \ba b$, $a \ba c$ the values related to $a$ need to be shared only once with the helper. In some cases, we might need two different encryptions. The need for two encryptions arises when evaluating circular structures, such as all three terms $a\ba b$, $a\ba c$ and $b\ba c$. The encrypted values and keys cannot be distributed such that the helper gets an encrypted value without getting the corresponding key to decrypt it. More precisely, to compute $a\ba b$ (see Figure \ref{fig:3and}), the helper gets the key $K_b$, the encrypted key $ENC_{K_6}(K_a)$ and the encrypted value $ENC_{K_a}(a)$. To compute $a\ba c$ the helper must get $K_c$ and the encrypted value $ENC_{K_c}(c)$, since it already received $ENC_{K_a}(a)$ and thus it cannot get key $K_a$.  To obtain $b\ba c$ is not possible, since the helper has already $K_b$ and $K_c$ and thus cannot get either $ENC_{K_c}(c)$ or $ENC_{K_b}(b)$.\\
To handle circular structures of the form above, two encryptions of the same variable suffice. Multiple encryptions of the same confidential variable can easily be created by the KH and EVH. To re-encrypt a variable $a$ encrypted with key $K_a$. The KH chooses a key $K_a'$ and transmits $ENC_{K_a'}(K_a)$ to the EVH, which computes $ENC_{ENC_{K_a'}(K_a)}(ENC_{K_a}(a)) = ENC_{K_a'}(a)$. It is not hard to see that for each variable out of $v$ variables we need to at most two encryptions to be able to compute any of the $O(v^2)$ possible pairs. For an AND involving variable $a$ we use the first encryption of $a$, if $a$ is on the left-hand side of the AND, e.g. for $a \wedge x$, and the second encryption, if it is on the right-hand side, e.g. $x \wedge a$.\\
Next, we discuss the adjusted protocols for the AND of two variables $a \ba b$ using three parties reusing priorly shared values with the helper. There are three cases: Reusing both variables, reusing one variable, reusing one variable and reencrypting the other. The protocol in Figure \ref{fig:3andBoth} shows the reuse of both operands of an AND. In this case, compared to the protocol without reuse (Figure \ref{fig:3and}) only two keys $K_7'$ and $K_8'$ need to be generated and shared as well as the value $ENC_{K_7'}(t_4)$.

\begin{figure*}[!htp]
	\centerline{\includegraphics[width=1.05\linewidth]{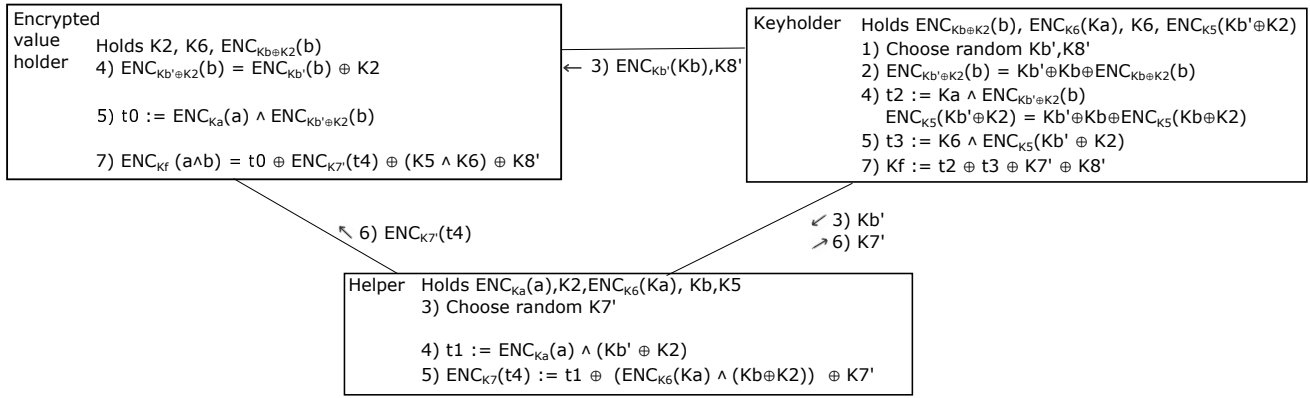}}
	\caption{Algorithm for an AND ($\ba$) operation of two bits $a,b$, reusing values for variable $a$ and re-encrypting variable $b$.}
	\label{fig:3andOne}
\end{figure*}

Reusing values for one variable $a$, while transmitting those of a variable $b$, works similarly. In the protocol in Figure \ref{fig:3and} two bits, ie. $ENC_{K_a}(a)$ and $ENC_{K_6}(K_a)$, do not have to be transmitted. 
Reusing values for one variable $a$, while reencrypting the other is shown in Figure  \ref{fig:3andOne}.
In this case, we first reencrypt one variable, i.e. $b$, and then use the reencrypted values. The KH sends $ENC_{K_b'}(K_b)$ to the KH. The EVH reencrypts $b$ to get $ENC_{K_b'}(b)$ and uses this value in his computations. Note, that the KH can reuse $ENC_{K_2 \opl K_b}(b)$ to get $ENC_{K_2 \opl K_b'}(b)$, i.e. the EVH and HE keep $K_2$ and the EVH does need to share $ENC_{K_b' \opl K_2}(b)$ to the HE. The same holds for $ENC_{K_5}(K_b \opl K_2)$.

\bth \label{thm:3andBoth}
The protocols in Figure \ref{fig:3andBoth} and \ref{fig:3andOne} are perfectly secure and correct.
\ethe
\bpr
Correctness follows from correctness of Figure \ref{fig:3and}, since all computations are identical. Security follows from the fact that no party obtains additional information. In the protocol of Figure \ref{fig:3andBoth} the helper obtains no input. The EVH obtains an encrypted value that was encrypted with a newly generated key $K_7'$. The KH only obtains the newly generated key $K_7'$.
In the protocol of Figure \ref{fig:3andOne} the EVH additionaly obtains the encrypted key $ENC_{K_b'}(K_b)$ and $K_8'$ but no information about $K_b'$. The helper obtains $K_b'$, but not $ENC_{K_b'}(K_b)$. The KH obtains $K_7'$ and $K_8'$.
\epr

\bth
For $v$ variables computing $t(v) \in [v,v^2/2]$ pairs, needs at most $1 + 4\cdot v/t(v)$ bits in total.
\ethe
\bpr
The protocol in Figure \ref{fig:3and} needs 5 bits to be communicated to compute an AND of two bits due to Theorem  \ref{thm:3bits}.
The protocol in Figure \ref{fig:3andOne} needs 2 bits using pre-shared keys, i.e. $ENC_{K7'}(t_4)$ and $ENC_{K_b'}(K_b)$.
The protocol in Figure \ref{fig:3andBoth} needs 1 bit, $ENC_{K7'}(t_4)$.
We need at most two encryptions for each variable. Sharing all parts of an encryption needs two bits at most, yielding $4v$ for sharing. Thus, to compute $t(v)$ terms, we need just one bit per term to return the result, giving a total of $4\cdot v + t(v)$ bits. 
\epr

%\begin{comment}
\subsection{Exponential Functions} \label{sec:exp}
We show how to compute the exponential function in the three party case ensuring statistical security. We assume that a secret $a$ is encrypted using additive blending without modulo of a random key $K$, i.e. $ENC_K(a)=a+K$. Our protocol computes $c^a$ for a public constant $c$ and a confidential value $a$. The protocol relies on the well known identity $c^{a+b}=c^a\cdot c^b$. 
The EVH shares a random key $K_1$ with the KH. It computes $ENC_{K_1}(c^{a+K})=c^{ENC_K(a)} +K_1$ and transmits this to the Helper.
The KH computes $K_3:=-K_1/c^K-K_2$ and transmits $K_3$ to the Helper.
The Helper receives $K$, $K_3$, $ENC_{K_1}(c^{a+K})$. It chooses a key $K_2$ and shares the key with the KH. It computes $ENC_{K_1}(c^{a+K})/c^K - K_3 = c^a+K_2 = ENC_{K_2}(e^a)$, which is shared with the EVH.
%\end{comment}

\subsection{Arbitrary Fan-in}\label{sec:arb}
% We discuss two ways. We can trade the number of required (synchronized) rounds to exchange messages for employing more entities. More precisely, to AND $w$ numbers rather than two numbers requires $2^w$ parties to get constant round complexity. This follows from Equation (\ref{eq99}) generalizing Equation (\ref{eqM}), i.e., there are $2^w$ terms with one term being computed by one entity. The second way uses only three parties.
We can compute $a_0 \wedge a_1 \wedge \ldots \wedge a_{w-1}$ using two rounds only. Similarly as for two variables, the AND of multiple variables can be expressed using multiple terms such that each term consists of ANDs of multiple encrypted values and multiple keys. The EVH can compute the AND of all encrypted values locally, and the KH can do the same for the keys. Thereby, reducing each term to one partial result held by the KH and one by EVH. These two partial results can then be ANDed using the prior protocol for two variables, yielding the AND of one term. The results of all terms are XORed. % of these two local terms%Each party encrypts its partial result, and then they perform an AND of two partial results to obtain the term. This can be done for all terms in parallel.

\bth
A gate $a_0 \wedge a_1 \wedge \ldots \wedge a_{w-1}$ can be evaluated in 2 rounds using messages of size O($2^{w}$) for an arbitrary parameter $k \in [2,w]$ and O($w 2^{w}$) bit operations.
\ethe
\bpr
We can express the AND using $2^w$ terms by generalizing Equation (\ref{eqM}). Let $S_{w}$ be all subsets of $\{0,1,...,w-1\}$. We have
% For a term consisting of $w$ variables, each variable $i$ can either be the key $K_i$ of secret $a_i$ or the encrypted value $ENC_{K_i}(a_i)$.

\begin{flalign} \label{eq99}
&a_0 \wedge a_1 \wedge \ldots \wedge a_{w-1}  \nonumber \\
= & (a_0 \opl K_0 \opl K_0) \wedge a_1 \wedge \ldots \wedge a_{w-1}  \nonumber \\
= &(ENC_{ K_0}(a_0) \wedge a_1 \wedge \ldots \wedge a_{w-1})  %\\ \nonumber
\opl  (K_0 \wedge a_1 \wedge \ldots \wedge a_{w-1})   \nonumber\\ 
= &(ENC_{ K_0}(a_0) \wedge (a_1 \opl K_1 \opl K_1) \wedge \ldots \wedge a_{w-1}) \nonumber\\ %\\ \nonumber
&\opl  (K_0 \wedge a_1 \wedge \ldots \wedge a_{w-1})   \nonumber\\ 
= &(ENC_{ K_0}(a_0) \wedge ENC_{ K_1}(a_1) \wedge \ldots \wedge a_{w-1}) \nonumber\\ %\\ \nonumber
&\opl  (ENC_{ K_0}(a_0) \wedge K_1 \wedge \ldots \wedge a_{w-1})   \nonumber\\ 
&\opl  (K_0 \wedge a_1 \wedge \ldots \wedge a_{w-1})   \nonumber\\ 
= &(ENC_{ K_0}(a_0) \wedge (ENC_{ K_1}(a_1)) \wedge \ldots \wedge a_{w-1})  \nonumber\\%\\ \nonumber
&\opl  (ENC_{ K_0}(a_0) \wedge K_1 \wedge \ldots \wedge a_{w-1})   \nonumber\\ 
&\opl  (K_0 \wedge (a_1 \opl K_1 \opl K_1) \wedge \ldots \wedge a_{w-1})   \nonumber\\
= &(ENC_{ K_0}(a_0) \wedge ENC_{ K_1}(a_1) \wedge \ldots \wedge a_{w-1})  \nonumber\\%\\ \nonumber
&\opl  (ENC_{ K_0}(a_0) \wedge K_1 \wedge \ldots \wedge a_{w-1})   \nonumber\\ 
&\opl  (K_0 \wedge ENC_{ K_1}(a_1)  \wedge \ldots \wedge a_{w-1})   \nonumber\\
&\opl  (K_0 \wedge K_1 \wedge \ldots \wedge a_{w-1})   \nonumber\\
= &\opl_{S_E \in S_{w}} ((\ba_{j\in S_E} ENC_{K_j}(a_j))%\\ &
\ba (\ba_{j\in \{0,1,...,w-1\}\setminus S_E} K_j))%\\  \nonumber
%=:& \opl_{l \in [0,2^w-1]} E'_l \opl K'_l$
\end{flalign}
%The last step in Equation (\ref{eq99}) is just a definition of $t_i$.
In the last  step  we applied to all $a_i$ the same transformation as for $a_0$ and $a_1$, i.e. replacing $a_i$ by $a_i \opl K_i \opl K_i$ followed by an expansion of terms. We rearranged using the commutative property of the AND operation. It can easily seen that each of the $w$ variables doubles the number of terms, yielding $2^w$ terms, i.e. each corresponding to one of the subsets $S_w$.
%For example,  \begin{multline}  a_0 \ba a_1 \ba a_2 = \\  ENC_{K_0}(a_0)\ba ENC_{K_1}(a_1)\ba ENC_{K_2}(a_2) \\  \opl ENC_{K_0}(a_0)\ba ENC_{K_1}(a_1)\ba K_2 \\  \opl ENC_{K_0}(a_0)\ba K_1\ba ENC_{K_2}(a_2)  \opl ... \opl K_0\ba K_1\ba K_2  \end{multline}

In Equation (\ref{eq99}) each term $t_i$ consists of ANDed values. It can be partitioned into two parts, one consisting of encrypted values $t_E$ and one of keys $t_K$, e.g. for $t=ENC_{K_0}(a_0)\ba ENC_{K_1}(a_1)\ba K_2$ we get $t_E=ENC_{K_0}(a_0)\ba ENC_{K_1}(a_1)$  and $t_K=K_2$.  The EVH can compute the term $t_E$ by computing the AND of all encrypted values without communication and the KH the term $t_K$ in the same manner. The EVH encrypts the locally computed term $t_E$ and the KH encrypts $t_K$, i.e. the EVH chooses key $K_{tE}$, computes $ENC_{K_{tE}}(t_E)$ and sends the key $K_{tE}$ to KH. The KH chooses $K_{tK}$, computes $ENC_{K_{tK}}(t_K)$ and sends the encrypted value $ENC_{K_{tK}}(t_K)$ to the EVH. Then they run the protocol (Figure \ref{fig:3and}) to AND the two terms $t_E \ba t_K$. They do this for all $2^w$ terms in parallel. Finally, the EVH computes the XOR of all encrypted results for all $2^w$ terms and the KH computes the XOR of all keys, which yields the final result for each party. %compute $\opl_{i \in [0,2^w-1]} t_i = \opl_{i \in [0,2^w-1]} t_{Ei} \ba t_{Ki}$. To compute $\opl_{i \in [0,2^w-1]} t_i$ can be computed using a constant number of rounds and messages of constant size.
\epr

\section{Related Work}
Helpers are not uncommon in MPC, e.g. \cite{bon97,du01}. But they are often used as trusted entities. In this work, we do not trust the helper more than any other party. In the setting of \cite{du01} a client wants to know if a value held by the party matches her secret string. A helper assists in answering the query. The result of the query should also remain secret to the party. Generation of RSA keys is discussed in \cite{bon97} using a helper. The helper is used to compute the product of primes using an interpolation of a quadratic polynomial. We integrate a helper on a much lower level of computation and adjust basic protocols like AND and XOR to use a helper. % . The product consists of shares

Though a large body of work \cite{Gol87,bea90,ara12,bar89,ben88,bog08} does not distinguish between encrypted values and keys, the idea of drawing such a separation has been employed in other contexts, eg. in the work of \cite{bog09} discussing MPC the idea of using such a separation with public keys for voting schemes is mentioned.

Three parties are commonly used, e.g. \cite{moh15,bog08,lau14}. The work \cite{moh15} builds upon garbled circuits, essentially showing that garbled circuits can be made robust against corruption of one party. Sharemind \cite{bog08} uses three parties and additive secret sharing, i.e. for a secret $x$ each party $P_i$ obtains a share $x_i$ such that $\sum x_i \mod 2^{32} = x$. To perform a multiplication they compute all 6 shares $x_i\cdot x_j$ using \cite{du01}. A multiplication requires 3 rounds and 27 messages each containing a 32-bit value. %An adaption of our scheme to multiplication would require at most 2 rounds and 4 messages.
The paper \cite{bog08} also discusses why Shamir's secret sharing fails on the ring of $2^{32}$ (and needs more messages on the ring $Z_p$). The work \cite{du01} also uses an untrusted third party, which assists in the computation of approximate distances (e.g. of strings) using various metrics. The system of \cite{lau14} uses three parties and linear secret sharing to compute all nine shares for evaluation of multiplication as \cite{bog08}. 
Recently, a new three party protocol\cite{ara12} was introduced. In contrast to this work (and similar to other works, e.g., \cite{bog08,mau06}), each party obtains a share using linear secret sharing. The protocol creates correlated randomness among all three parties. We use correlated randomness for pairs of parties. They do not require a round for secret sharing as we do, but assuming that each variable appears on average (somewhat) more than 10 times in a circuit, their protocol \cite{ara12} requires more communication to evaluate the circuit. In particular, if the number of terms $t(v)$ is more than linear as the number of variables, we need only $2+o(1)$ bits per gate and thus outperform by a factor of 3.\\

An unbounded fan-in AND gate can be simulated \cite{bar89} in (expected) constant number of rounds for arithmetic gates. They encrypt a number $a_i$ held by party $i$ as $ENC(a_i)=R_{i}\cdot a_i\cdot R_{i-1}^{-1}$ with $R_i$ being a matrix of random elements. The product of all terms $a_i$ is one element in the matrix being the product of all encryptions. To generate matrices of sufficient rank, they generate more than $n^2$ random matrices. We do not use multiplicative inverses in a group. We follow a different approach based on term expansions. Bar-Ilan et al. \cite{bar89} requires messages that are of size proportional to the size of a constant depth, unbounded fan-in circuit for the function to evaluate. Our scheme (JOS) requires asymptotically also a constant number of rounds for computation of a $w$ fan-in gate but more communication for large $w$. For the three party case, JOS outperforms \cite{bar89} for small fan-in gates. For example, for $w=4$ Bar-Ilan et al. requires at least 6 rounds (using $k=3$), whereas JOS needs at most five. The total amount of communication of \cite{bar89} is at least $129 \cdot l$ in contrast to $60\cdot l$ of our scheme. Furthermore, it needs more local computation. % and is targeted towards Boolean circuits

% our: mul (6*3*l + 2* 6l) * 2
% $w\cdot l \cdot 2^{w-k}$: k=4 :  $16\cdot l$,
%bar89 rand matrix generate: $13 \cdot 9 \cdot p$, >= 6 rounds; multi of inv and sec and inv 12*l  => 129*l

%that takes only 1 round and 6 messages in a three party setting.  Through pre-sharing of keys, we also require 1 round but only 2 messages (without pre-sharing 4 messages) for Boolean Circuits. For arithmetic circuits, we need 2 rounds

The BenOr-Goldwasser-Wigderson (BGW) \cite{ben88,ash11} gives several fundamental MPC protocols. %paper states aside from giving several MPC protocols that $t$ private protocols allowing for collusion of $t\geq n/2$ are impossible. The proof discusses only the case $n=2$ for the OR function, stating that some information leakage takes place when the OR function is computed (and revealed). We only reveal results to the client. We can tolerate collusion of up to $n-2$ parties (for $n\geq 3$).
Genaro-Rabin-Rabin (GRR) \cite{Gen98} simplifies BGW. GRR requires $n\geq 2\cdot t+1\geq 3$ parties tolerating collusion of $t$ parties, BGW can handle collusion of $t<n/3$ parties in the distrust all model.  GRR and BGW use Shamir's secret sharing to derive a protocol for multiplication. The multiplication protocol Simple-Mult in GRR takes two secrets $\alpha$ and $\beta$ shared by two polynomials $f_{\alpha}(x)$ and $f_{\beta}(x)$ to compute $\alpha\cdot\beta$. Party $i$ computes the value $f_{\alpha}(i)\cdot f_{\beta}(i)$ using a random polynomial. Then each party aggregates the input of other parties and reduces the size of the polynomial through interpolation to compute his share of $\alpha\cdot \beta$.% The complexity of GRR is higher than an adaptation of our AND protocol to multiplication asymptotically, but with larger constants and more complex operations:
A protocol for multiplication and addition using similar ideas as GRR but using additive secret sharing (without modulo) is given in \cite{mau06}. In the case of three parties a secret $a$ is split into three parts $a_0,a_1,a_2$ such that the sum equals $a$. In \cite{mau06} each party gets two distinct parts. Multiplication of two secrets $a$ and $b$ is analogous to GRR by computing all nine pairs $a_i\cdot b_j$, aggregating them locally and sharing the result using independent randomness. A party then aggregates all received numbers to obtain the result $a\cdot b$. To compute $(a\cdot b) \cdot c$ each party would send its share of $a\cdot b$ to one other party, such that each party again holds two shares of the result. A key disadvantage of \cite{mau06} is that shares double in size after every multiplication, making it impractical for even a modest number of multiplications. 
The paper by Yao \cite{yao86} from the late 80ies still forms the underpinning for many works evaluating Boolean circuits. Yao showed how one party $A$ can evaluate a private boolean circuit with private inputs from itself and another party $B$ such that $A$ does not learn anything about the inputs of $B$ and $B$ does not learn anything about the circuit or the input of $A$. To do so $A$ computes a so called ``garbled circuit'' which is an encryption of the circuit containing the input. Afterwards, party $B$ evaluates the encrypted circuit using its input and returns the result. Encryption encompasses encrypting every entry of the truth table of the boolean circuit and uses several algorithmic ideas such as oblivious transfer of keys to do the two party computation. The original scheme \cite{yao86} allowed for a circuit only to be evaluated once without revealing information about the circuit. Since then a lot of improvements have been made, e.g., \cite{gar13,Gent13,bell13}. Reusable circuits come only with additive overhead in the form of a polynomial in the security parameter and circuit depth \cite{Gent13}. Our advantage compared to \cite{Gent13} is that we ensure perfect security and encryption is much simpler (and faster). Additionally, our communication complexity does not depend on a polynomial depending on the security parameter as well as the circuit depth, which can easily dominate the communication costs. Yao's scheme has been generalized to multiple parties by computing a common garbled circuit in BMR \cite{bea90}. Several evaluations are possible using multilinear jigsaw puzzles \cite{gar13}. A circuit can be garbled such that its encryption occurs only additional overhead \cite{Gent13}. Recent implemenations\cite{bell13} allow for a single AES call per garbled-gate (justified in the random-permutation model).
Goldreich-Micali-Widgerson (GMW) \cite{Gol87} uses oblivious transfer to compute any Boolean circuit. Values are encrypted such that each party holds parts of the non-encrypted value. The GMW protocol has round complexity linear in the depth of the circuit. Oblivious transfer has been continuously optimized, e.g. \cite{Ash13} uses symmetric cryptography. Still, using \cite{Ash13} for an oblivious transfer requires (as a lower bound) at least the size of the security parameter, which is significantly more than our total communication for an $AND$.

A significant body of work has focused on optimizing either the computational or communication overhead (e.g. \cite{dam10,dam13,dam07,ish09}) of MPC focusing on entire circuits for various security models using known schemes for evaluating gates. We focus on optimizing elementary operations for a single gate for perfect security that can be used to compute entire circuits.
There is a vast number of secret sharing schemes, e.g. for a survey see \cite{bei11}. Our linear encryption schemes are known. For instance, \cite{ito89} encrypts a secret using XOR. Additive encryption as done in JOS roughly corresponds to \cite{ben88} and has been also employed by \cite{Cat10}. Whereas prior work shared a secret with all parties, we use a dedicated helper to support computation and use the properties of the encryption schemes to derive novel protocols.

The first work attempting to compute exponential functions in a constant number of rounds was \cite{luo04}. The model of \cite{luo04} assumes that each party has a secret. Although they claim that an adversary can corrupt two parties for all their protocols, in fact, information about a secret (of one party) is revealed if one party behaves dishonestly. To see this consider, e.g. the protocol 3.1 for multiplication in \cite{luo04}. They compute $u_1+u_2=x_1\cdot x_2$ using a non-specified algorithm, where $x_1,x_2$ are secrets, $u_i$ are secret shares and $u_i,x_i$ are held by party $i$. If an attacker corrupts parties 1 and 2 and obtains $u_1$ and $u_2$ it also obtains $x_1\cdot x_2$. Clearly, revealing $x_1\cdot x_2$ is a violation of confidentiality for both $x_1$ and $x_2$. Furthermore, they require a two-party protocol (as blackbox) for real numbers to get $u_1,u_2$ such that $u_1+u_2=x_1\cdot x_2$ but do not state any efficient protocol. %Even if a two-party protocol for multiplication of reals existed with the same efficiency as existing multiplication protocols for integers, their protocol would require more rounds: They require two multiplications each consisting of three shares, i.e. $x_0\cdot x_1 \cdot x_2$. Using 32bit integers this could be done using \cite{bar89} or by computing $r=x_0\cdot x_1$ and then $r\cdot x_2$ \cite{ara12}. The later requires a total of four rounds and 4 multiplications each using 32 bits integer requiring 384 bits. An adaption of our AND protocol to multiplication would require two rounds, 4 messages and the transmission of 160 bits, i.e. 5 numbers each having 32 bits.
%The first work to compute exponential functions in a constant number of rounds was \cite{dam06}. In fact, they discuss the case to compute $y^x$ with private $y$ and for public and private $x$. Note, we compute any exponential function $y^x$ where $y$ is public, e.g. covering the standard exponential function $e^x$ with the Euler constant $e=2.718...$. \cite{dam06} requires computation of inverses and comparison to zero. \cite{dam06} was improved by \cite{nin11}, which uses several ideas from \cite{dam06}. \cite{nin11} need more than 15 rounds and more than $157\cdot l$ bits for $l$ bit numbers. Our computation takes two rounds and xy to get stat sec z
This work is an extension of \cite{sch15ab} with more detailed model discusion, proofs, computation of the exponential function, a protocol for reusing values and a protocol for achieving a trade-off between round and communication complexity for large fan-in gates. Prior work \cite{sch15b} based on \cite{sch15ab} has shown how to compute various logical and numerical functions, e.g. trigonometric functions and divisions, as well as transformations between different encryption schemes.

\section{Conclusions}
We have assessed the idea of using one party as a helper in the context of secure-multi party computation. The derived protocols achieve little communication, storage, and computational overhead. In some cases, they are theoretically optimal regarding communication complexity, showing that a helper can be of great value from a theoretical perspective. Numerical computations relying on statistical security can in some cases also largely benefit from a helper. % This makes it easily implementable and a suitable candidate for a framework targeted towards privacy preserving data mining, where a minimal amount of communication as provided by JOS is essential. Thus, in future work, we want to implement an industrial strength system focusing on numerical operations.

\bibliographystyle{abbrv}
\bibliography{refs}

\end{document}